# Reconsidering Mermin's "In praise of Measurement"

GianCarlo Ghirardi
Department of Theoretical Physics, University of Trieste,
The Abdus Salam ICTP, Trieste,
and
Istituto nazionale di Fisica Nucleare, Sezione di Trieste, Italy

**Abstract**: We critically analyze a recent paper by D. Mermin and we compare his statements with Bell's position on the problems he is discussing.

## 1. Introductory Considerations.

Recently, an interesting paper by D. Mermin [1] appeared, by the (deliberately) provocative title *In praise of Measurement*, which, for people interested in foundational issues, acquires its whole significance when confronted with one of the last and probably the most well known of the foundational papers by the late J.S. Bell [2]: *Against Measurement*. The paper is in line with the recent positions taken by the author who, to use his words [3], in spite of the fact that he *was entirely on Bell's side on the matter of knowledge-information,* subsequently *fell into bad company and started hanging out with the quantum computation crowd, for many of whom quantum mechanics is self-evidently and unproblematically all about information … .* In this paper I pay attention to some aspects of the paper which, in my opinion, deserve a comment, among them those points in which the author makes explicit reference to Bell's views and strict requirements concerning quantum mechanics.

## 2. The measurement gate.

The paper attaches a very particular importance to the 1-Qbit measurement gate. I completely share the author's opinion that such a gate deserves a specific attention, not only for the extremely important role it plays for the actual implementation of quantum circuits, but for the fact that it is the only gate whose output is not uniquely determined by the input and whose action is not accounted by a unitary transformation. I consider also extremely appropriate Mermin's stressing that measurement gates play an absolutely fundamental role in the preparation of a state (the inputting process) and that they are an essential ingredient of the error correction procedures.

From my point of view (which I imagine is not shared by Mermin) these facts are of the utmost importance because they compel people to realize that the superposition principle (i.e. the linear character of the equations governing natural processes) has to be given up, at an appropriate level, even in order to account for the common everyday laboratory practice. This conviction has led A. Bassi and myself to present [4] '*A general argument against the universal validity of the superposition principle*' showing how elementary (and in our opinion irrenounceable) requirements concerning the fact that we can get information about a microsystem by leaving it interact with a macrosystem in such a way to lead to perceptably different macroscopic final situations, impose to accept the conclusion of von Neumann that, at a certain level, one has to give up the linear structure of the theory, one has to take into account that in nature nonlinear processes must occur. Our proof does not make recourse to any of the restrictive assumptions of the original argument as well as of those used by many researchers, a typical example being the extremely refined analysis presented by B. d'Espagnat in his famous book [5]. Actually, d'Espagnat himself has recently claimed [6] that in ref. [4] *'the impossibility of reconciling the idea that the measurement process is a purely physical phenomenon with standard quantum rules was indeed proved, and in a very elegant and, I think, truly general manner'*.

I hope to have made clear while I welcome the particular emphasis that D. Mermin has given to the essential role of the 1-Qbit measurement gate, a device which violates the linear nature of the theory.

I must also say that I fully share the thesis of Mermin's paper that, by simpy considering – in addition to the unitary boxes whose functioning is well accounted for by the unitary quantum evolution - non-unitary boxes which work according to the wave packet reduction postulate, one can safely go on in assembling useful devices implementing efficient algorithms. I will come back to this point in what follows.

Let me discuss now the points of Mermin's paper which involve directly Bell's views.

### 3. A preliminary remark.

I must confess that I have some difficulty in grasping the purpose of Mermin's comments concerning Bell's views on the foundational issues of quantum mechanics and his hypothetical reactions would he have been exposed to the quantum-computation revolution. Even though Mermin avoids to commit himself to what would have been Bell's position, he seems to think that the new scenario and, in particular the consideration of the precise role and virtues of the 1-Qbit measurement gate, might have weakened or even cancelled Bell's concerns with quantum theory so clearly expressed in *Against Measurement*. It is difficult for me to take this point seriously. In fact, since all of quantum computation is just and plainly an application of the quantum formalism, I do not see any reason whatsoever which might make it (a) superior to the quantum formalism itself or (b) of any use in overcoming the difficulties that physicists have met with it, in particular those which are at the basis of John Bell's unsatisfaction with the theory.

I hope that the ensuing considerations and the reference to precise statements by Bell, will make more clear the reasons for my difficulties with part of ref. [1].

### 4. Exact Theories.

D. Mermin makes a precise reference to Bell's requirement that a theory be "exact". To illustrate this fact he mentions only the feature which Bell, in *Against Measurement* has claimed should characterize an exact theory, i.e. that it is *fully formulated in mathematical terms, with nothing left to the discretion of the theoretical physicist*. And it is with reference to this formulation that Mermin concludes that, *since all of the gates* [of quantum computation], *including the measurement gate, alter the state associated with the incoming Qbits in a well defined, generally discontinuous manner, which is precisely defined by that state, with nothing left to the discretion of the theoretical physicist*, such a theory meets Bell's requirement of exactness. However, I think that, in this way of presenting the situation, the author misleads the reader concerning Bell's pretension of exactness, because he does not mention the many other occasions (which, in my opinion, render better his views on this point) in which he has made precise his idea of exactness. In particular, to illustrate this point it is useful to make reference to a sentence (among many) in which Bell has made clear [7] why he considered Classical Mechanics a *theory which is not intrinsically inexact*, the fundamental reason being that *it neither needs nor is embarassed by an observer*. And he has stressed this point again and again in his writings, up to the very last one. I have discussed this matter many times with him personally, and I have no doubts that he considered this as the basic feature of an exact theory.

Now, Mermin, after having brilliantly stressed how one can play with the 1-Qbit measurement gate by resorting to one or to an arbitrary number of ancillas (the analogous of the shifty split of the standard theory) and having claimed that now there is nothing shifty in the game because the crucial role is always played by the 1-Qbit measurement gate, states that *it is through the readings of 1-Qbit measurement gates, and only through such readings, that one can extract information from a quantum computer*. And, even after the subtle and smart analysis concerning the preparation procedure, the conclusion does not change: *we have therefore failed in our attempt to eliminate measurement gates from the act of state preparation* and/or of *error correction*. Even more: *measurement is essential at both ends of a quantum computation*. However, according to Mermin, his position concerning quantum information overcomes Bell's criticisms, because: *in spite of the fact that there remains the distinction between linear, invertible, unitary gates with output state fully determined by the input, and nonlinear, irreversible, measurement gates, with output state stochastically determined by the input,… the action of both types of gates are fully defined, with nothing left to the discretion of the theoretical physicist*.

Once more I think that this sentence is seriously misleading. Actually, if one reads carefully Bell's papers, one grasp immediately that the above remarks in no way might have cancelled his worries with the theory. In fact, the above sentence asserts that there are two fundamentally different types of gates and claims that the theoretical physicist knows clearly how to deal with each one, but the author does not present any precise physical criterion to characterize the two classes. The distinction is assumed to be easy and unambiguous for the *theoretical physicist*. To analyze this aspect I feel the need to quote another important and sharp sentence [8] by Bell concerning the standard quantum approach: *There is nothing in the mathematics to tell what is "system" and what is "apparatus", nothing to tell which natural processes have the special status of "measurements". Discretion and good taste, born of experience, allows us to use quantum theory with marvelous success, despite the ambiguity of the concepts named above in quotation marks. But it seems clear that in a serious fundamental formulation such concepts must be excluded*.

It goes without saying that I, like Bell, with reference to Mermin's counterpart of the expressions "system, apparatus, measurement", i.e. the unitary and 1-Qbit measurement gates, am perfectly aware that *discretion and good taste* play a fundamental role for the practical use of the quantum theory, but this does not eliminate the unsatisfaction with the structure of the theory, in its original form and/or in the quantum information version.

Concluding on this first issue, I claim that the quantum information approach, as outlined by Mermin, does not meet Bell's requirement. It seems to me that it is precisely to avoid this inescapable conclusion that in his paper Mermin has reduced quantum computation to a sort of game which one can play by using unitary and nonunitary boxes. It should be obvious that if one assumes that such boxes can be precisely and unambiguously identified and, moreover, their functioning must be left unanalyzed, one can avoid facing the puzzling aspects of the measurement box. I will come back to this point later, for the moment I will pass to consider the second of Bell's requirements.

5. **The requirement of a theory being serious**.

D. Mermin recalls that Bell has qualified as serious a theory which *should cover some substantial part of physics*, and also such that in it *apparatus should not be separated off from the rest of the world into black boxes*. Concerning the second part of the above statement Mermin admits that in quantum computation the elementary components of a circuit are dealt with as if they were black boxes, but he also points out that the same practice (unproblematically) characterizes classical computer science. I do not believe that the same practice can be unproblematically followed in the quantum context, and I will return on this point in the next section. Before doing this, I would like to concentrate my attention on the first part of Bell's sentence. When making statements about scientific schemes covering a substantial part of physics, Bell had always in mind and was making explicit reference to the explanatory character of big theoretical constructions such as Classical Mechanics, Maxwell Theory of Electromagnetism, Quantum Mechanics and so on. In his paper, Mermin - even though admitting that he is not sure that Bell would have viewed quantum computation as a serious part of quantum mechanics - claims that *any fragment of physics, large enough to be applied to the efficient factoring of enormous integers, has to be viewed as substantial*. And he adds: *the U.S. National Secutity Agency surely does*. No doubt this is so, but what should seem substantial to the National Security Agency for its purposes has little to do with what would seem substantial to a theoretical physicist who, like Bell, is concerned with the the laws of nature at a fundamental level.

Now, in order not to be misunderstood, I want to make clear that I consider extremely relevant and interesting, a real brainwave, the elaboration of Shor's algorithm [9]. But to claim that the *efficient factoring of enormous integers* has to be regarded as a substantial part of physics, is a position I cannot share for two reasons.

First, even if one is keen to accept Mermin's position concerning the factoring problem, one cannot forget that it is not what you can get, from a practical point of view, from a formal set of recipes which makes legitimate to qualify them as a serious theory. For instance, nobody would deny that a model which accounts for the whole family of spectral lines of the hydrogen atom covers a substantial part of physics, but, in spite of this, everybody would agree that Bohr's theory is internally inconsistent and requires a radical change in order to be taken as a serious theory.

Secondly, with reference to the point of the practical success of a theoretical scheme I consider important to stress that Bell has always recognized the absolutely exceptional predictive power of quantum mechanics, but that, in spite of his clear awareness of this fact, he always insisted that Quantum Mechanics is perfectly appropriate **only** FAPP (*for all practical purposes*). And it is this basic feature of the FAPP value of the theory which worried him and it is precisely the same fact that, I believe, would have not changed of a bit his critical position about the theory even if he would have witnessed the quantum computation revolution.

**6. The problem of the black boxes.**

I will reconsider here the problem of the legitimacy of separating part of the world from the rest by making recourse to black boxes, the fundamental procedure which stays at the basis of Mermin's analysis. I perfectly agree that, in principle, and for the limited purpose of assembling a circuit made up of many components, one can simplify his approach to, e.g., classical computation, by considering the various constituents of the whole circuit as black boxes, each of them performing a precise task. So, for an engineer, the only interest he can have concerning the various parts of his assemblage is to make precise the various outputs they yield when triggered by different inputs.

I believe, however, that, even at the level of classical computation, there is another aspect at least as important as the one we have just discussed that must be taken into account. In fact since, after all, as Mermin recalls continuously, it is only the user who can be interested in resorting to a computer for his purposes, there must be somebody (who may very well be the potential user himself) who is able to build up the black boxes yielding their characteristic input-output interplay. Now, how does this black-box builder works? He has a knowledge of a theory (e.g. classical electromagnetism) which tells him how he has to combine switches, wires, and so on so that, according to the known and precise natural laws which govern the physical processes he is taking advantage of, the final result of his action will actually be a black box working as desired. I want to stress that the theory he is using might be claimed to "explain" why, operating in this way, he can reach his aim.

The same argument holds, I believe, for the guy who builds the unitary constituents of a complicated quantum computer. In order to get black boxes working as desired by resorting to precise carriers of information (e.g., photons), he has to take into account the quantum behaviour of the physical components which go in a box. If he wants to have a box working, e.g. as the C-Not box, he will use appropriate optical devices, such that the behaviour of the photons, when imputted into these devices, will be the one predicted by the theory which he has at his disposal and which works perfectly well at this level: quantum mechanics. In this sense he understands completely how the unitary boxes work, he has full control of them, the theory makes absolutely clear how and why they bring about the desired result. Even more, it is his knowledge of the theory that guides him in actually building up the specific unitary boxes he will use.

The same cannot be said concerning the nonunitary measurement box, this absolutely essential element of any quantum circuit. In such a case, even though one can know what he has to insert at a certain stage of the device (e.g. something leading to a displacement of a macroscopic part of the system) in order to get the desired stochastic reduction of the wave packet, one has a quite vague knowledge of the details of the process, and actually one is perfectly aware that something in that box is surely not governed by the rules of the theory he is applying for the rest of the circuit and by the laws that scientists have identified as governing natural phenomena.

Is not this situation quite peculiar? Is it not another instance of the situation according to which one knows that at a certain level the superposition principle, this milestone of quantum mechanics, has to be abandoned, but he has not any idea of why it is so, and, even more, which is the precise level at which he has to forget the extremely successful theory he is using, to give up the deterministic and unitary evolution laws and to replace them by stochastic and nonlinear purely pragmatic recipes?

I am perfectly aware that I am raising here a problem which many physicists, in particular those involved in quantum computation, would consider as a problem which is not scientifically relevant. But is it really so? To conclude this paper let me spend some time in considering this last question.

**7. Physical versus metaphysical issues.**

The debate on quantum mechanics has seen various changes in recent years. First, important papers like the one by Joos and Zeh [10], have led to what J. Bub has qualified [11] as *the new orthodoxy*. It claims that the measurement problem is a pseudoproblem which finds its natural (dis)solution when one takes into account the unavoidable and uncontrollable degrees of freedom of the environment which must be traced out. It goes without saying that this solution must necessarily make reference to the ensemble rather than the individual level of description of physical systems. And this fact raises some serious problems (not so often presented in the open) also for the adherents to this position. For instance, in ref. [10] one reads: *The local description "is assumed" and the specific choice of a basis can **perhaps** be justified by a fundamental **inderivable** assumption about the local nature of the **observer**…* (the emphasis is mine).

Then came the (perfectly justified and important) interest in quantum computation brought out by the possibility of working directly with Qbits. Note that quantum computation unavoidably deals with individual rather than with ensembles of physical systems.

Thus, the new perspective accompaigning the new promising developments has given rise to what I will call *the newest orthodoxy*: quantum mechanics is not about the world, it is exclusively about information. I remeber that, even before the death of J. Bell, the suggestion of looking at the theory from such a perspective has been put forward in various occasions, in particular with reference to the IGUSES characterizing some of the variants of the decoherent histories approach. The reaction of Bell was sharp and precise. He refused to consider such a position unless [2], in advance, one would have answered to two basic (for him) questions: *whose information?*, and: *information about what?*

I know very well that Mermin disagrees with this position of Bell. In fact, in ref. [3] he stated that "Information about what?" *is a fundamentally metaphysical question that ought not to distract though-minded physicists*. Let me consider this position which seems to me to be in perfect consonance with the requirement of considering the measurement box as a black box. What matters is how a device works, not what are the precise laws that account for the way it works. In absence of any precise theory which allows one to "look inside" the nonunitary box, just as quantum theory allows to open the unitary ones, this position might be accepted (even with some uneasiness).

But the present situation is quite different. Anybody would agree that the nonunitary boxes are always unavoidably related to an interaction of a microsystem with a macrosystem. Now, there are two kinds of *exact* theories, hidden variable theories [12] and dynamical reduction models [13] – and Bell himself has used [2] the expression exact to refer to them - which give a fully satisfactory description of what happens in such situations. Moreover, while those of the first kind (in particular Bohmian Mechanics) have implications agreeing fully with the standard theory, those of the second (in particular the Spontaneous Localization Theories) qualify themselves as rival theories which can be explicitly tested against quantum mechanics. I would like to stress that even if they would turn out to be wrong they might very well give some precise hints about where the breakdown of the superposition principle might occur. It is my conviction that to ignore that the macro-objectification process (the nonunitary evolution under measurement) admits an empirical solution is a scientifically unappropriate position which amounts to embarque onself into a degenerate research program, in the terminology of I. Lakatos. This is clearly exemplified by those attempts which, to save the linear and unitary character of the theory, derive an equation of the Quantum Dynamical Semigroup type for the statistical operator mimicking the one of dynamical reduction theories from a fictitious theory formulated by resorting to an Hilbert space enlarged by the consideration of an undetectable hypothetical ancilla. This point has been analyzed with great clarity by A. Hagar in ref. [14].

Concluding, I think that the newest ontology is not making a good service to physics. A much more scientifically serious attitude is the one of people like S. Adler [15] and R. Penrose [16], who are trying to identify, within quite different conceptual frameworks, whether, where and when the linear nature of quantum mechanics might break down. This program, if brought to a consistent end and if it would lead to identify violations of quantum mechanics, would attribute to the important 1-Qbit measurement gate a completely different role as the prototype of those devices which would correspond to the new theoretical perspectives.

**8. Conclusions**

My position should be clear. I welcome any suggestion which points out clearly that the superposition principle must be violated at an appropriate level. For these reason I have appreciated a lot the paper by Mermin which attaches an extreme importance to devices characterized by a nonunitary dynamics. I also agree that, for the limited purposes of the quantum information practice, the non unitary box can be left untouched and taken as a black box, the only property of which is the one which is encoded in the wave packet reduction postulate and in the associated probabilistic assignements. But I mantain that I consider inappropriate to avoid facing the problem of 'looking inside' such boxes, particularly in view of the fact that some (even small) steps in accounting for their dynamical behaviour have been made and, at least, they have thrown some light on processes which might put into evidence the necessity of working out a more general theory than quantum mechanics itself. I conclude by claiming that, from the point of view of the scientific enterprise, to go on trying to clarify, to account in a precise way for the nonunitary nature of the boxes inducing wave packet reduction, is a must and that to leave the 1-Qbit box unanalyzed amounts to give up the search for a more satisfactory and exhaustive understanding of natural processes.

**Acknowledgments.**

I am indebted to Profs. S. Goldstein and R. Tumulka for extremely useful discussions and remarks.

**References.**